# Pulse Shaping of Broadband Adiabatic SHG in Ti-Sapphire Oscillator

Assaf Levanon[1], Asaf Dahan[1], Achiya Nagler[1], Erga Lifshitz[1], Eyal Bahar[1], Michael Mrejen[1], Haim Suchowski[1,2*]

[1]Condensed Matter Physics Department, School of Physics and Astronomy, Tel Aviv University, Tel Aviv, 69978, Israel
[2]The center of Light-Matter Interaction, Tel Aviv University, Tel Aviv, 69978, Israel.
*Corresponding author: haimsu@post.tau.ac.il

**Abstract**

**We experimentally demonstrate an efficient broadband second harmonic generation process with tunable mode-locked Ti: sapphire oscillator. We have achieved a robust broadband and efficient flat conversion of more than 35nm wavelength, by designing an adiabatic aperiodically poled potassium titanyl phosphate (KTP) crystal. Moreover, we have shown that with such efficient flat conversion, we can shape and control broadband second harmonic pulses. More specifically, we assign spectral-phase of abs-value and $\pi$-step, which allows wavelength tunable intense pump-probe and amplitude modulation of the broadband second harmonic output. Such spectral phases serve as a proof of concept for other pulse-shaping application for nonlinear spectroscopy and imaging.**

Ultrafast laser sources are in the heart of ultrafast experimental science. In the past two decades, ultrashort pulse lasers oscillators and amplifiers became well common equipment in the fundamental scientific exploration as well as in handful of industrial applications. Those sources, which by their nature are broadband and coherent, allows exploring many phenomena that occurs at the ultrafast time scale of many scientific processes and dynamical evaluations in nature [1-3]. Due to their extremely high peak power, nonlinear optics in the ultrashort regime results in an enhanced efficient frequency conversion generation processes, and is therefore of great interest in vast number of fields, such as color generation, nonlinear spectroscopy, imaging for metallurgy, photo-induced dynamics, noninvasive background free diagnostics and the generation of new color sources [4-6]. As these ultrashort pulses are much faster than any electronic system, many unique methods have been developed in recent years in order to characterize those pulses. Among them one can find ultrafast pump-probe and interferometric capabilities, allowing femtosecond temporal resolution experiments and ultrafast characterization methods, such as Frequency resolved optical gating (FROG), Multiphoton intrapulse interference phase scan (MIIPS) and spectral phase interferometry for direct electric-field reconstruction (SPIDER) [1, 7-9]. Among the various nonlinear conversion processes, three wave mixing and especially second harmonic generation (SHG) became widely used. Yet, frequency conversion of the ultrashort regime remained quite complicated, as the conventional conversion device usually exhibit a tradeoff between the conversion bandwidth and the conversion efficiency, rooted in the phase mismatch between the interacting waves, which is usually compensated only for a narrow band of frequencies [10, 11].

In the last decade, adiabatic frequency conversions have gained high interest and has been the subject of vast theoretical and experimental research [11-18]. The suggested adiabatic method has shown to overcome the tradeoff between conversion efficiency and bandwidth for sum frequency generation (SFG), difference frequency generation (DFG), Optical parametric amplification (OPA) and recently in SHG processes. In the fully nonlinear regime, where all interacting waves considerably change their power, efficient broadband conversion is even more difficult due to the complex nonlinear dynamics. In the past few years, there were a tremendous theoretical and experimental efforts to reconcile the requirements of the fully nonlinear regime processes and the high efficiency broad-bandwidth frequency conversion using the adiabatically varying design. A comprehensive theory for an adiabatic frequency conversion process of any quasi-CW three waves was developed by Porat and Adie [19] and by Phillips et. al. [20], later validated by Leshem et. al. [21] for the case of adiabatic SHG in the nano-second regime. The theoretical generalization and the experimental demonstration of adiabatic SHG for ultrashort pulses, which also take into account dispersion effect and higher order nonlinear parasitic effects, was later exhibited by Dahan et. al. [22], demonstrating an efficient robust frequency doubling for 75nm acceptance bandwidth, thermal acceptance of more than 100ºC and chirp variation of 300fs-3.5ps, thus displaying an unmatched robustness under both environmental conditions and characteristics of the incoming pulse. Though formerly demonstrated achievements of adiabatic frequency conversion in various nonlinear processes, all have used amplifiers' pump energies for satisfying the adiabatic criteria.

In this letter we experimentally demonstrate that adiabatic design is capable of an extremely robust efficient SHG also in conventional femtosecond high rep-rate oscillators powers. We show that with pulse peak energies of nJ-regime, one can achieve above 50% of energy conversion efficiencies for 70fs Ti-Sappire pulses. Furthermore, the flat conversion bandwidth response of the presented design, allows to perform broadband pulse shaping manipulations prior to the nonlinear optical conversion, thus not suffering from the spectral limitation that conventionally imposed by the limited bandwidth of birefringence or regular periodic crystals designs. More specifically, using spatial light modulator (SLM) in a 4-f pulse shapers, we present the implementation of a tunable pump probe apparatus based on a varying absolute phase spectral shape function in the frequency domain. In similar way, we show that when applying a π-step spectral phase, a coherent control of the SHG spectrum can be achieved, imposing a complete dip in the SHG, which is an outcome of complete destructive interference between the convolution of the interacting fundamental waves [23].

The experimental setup, illustrated in Fig. 1, consists of a 80 MHz repetition rate tunable coherent oscillator between 690nm to 1040nm (Mai-Tai) which served as the pump pulses, delivering 17.5nm full width half maximum (FWHM) ~70fs transform limited pulses, with energies spanning between 10-30nJ. The pump pulse then passes through JENOPTIK 640D spatial light modulator (SLM) in the range of 430-1600nm, enabling us to alter the pump pulse spectral phase and temporal behavior before it is focused into an adiabatically aperiodically poled KTP (adAPKTP) crystal. The residual pump and the generated SHG were separated using a dichroic mirror around 950nm. The spectrum of the SHG pulse was measured using Avantes spectrometer and is displayed in the right lower image of Fig. 1, thus demonstrating an efficient conversion efficiency all over the pump spectra.

The phase mismatch between the interacting waves was compensated in the poling method [10]. In the presented design, the second order nonlinear susceptibility χ(z) fluctuates between +χ to –χ in batches determined by $\chi(z) = \text{sign}(\cos(K_g(z)z))$, where the grating function $K_g(z)$ is given by

$$K_g(z) = (118.2 z_{cm}^3 - 45.2 z_{cm}^2 - 997.9 z_{cm} + 7957.1)[cm^{-1}]$$, and z spans all along the crystal length $z \in [-5mm, 5mm]$.

In our analysis, we used a three-dimensional generalization of the fully nonlinear dynamical equations dictating nonlinear conversion of any three wave mixing and in particular the depleted ultrashort SHG [22]. Two photon absorption of the fundamental pump and the generated SHG was also taken into account as in case of Ref. [22], likewise found to be of great importance. Since the SHG process occurred within the Rayleigh range of the pump pulse, spatial diffraction can be neglected, thus eliminating interaction between different areas of the transverse intensity profile.

The following three-dimensional generalization of the SHG equations (1) and (2), were therefore applied for predicting the SHG process behavior:

$$\frac{\partial B_{SHG}(r,z,t)}{\partial z} + i\mathcal{F}^{-1}\big(\beta(\omega + \omega_{SHG}) B_{SHG}(r,z,\omega)\big) = -i\chi(z)\mathcal{F}^{-1}\left(\frac{\omega + \omega_{SHG}}{n(\omega + \omega_{SHG})c} \mathcal{F}\left(B_p^2(r,z,t)\right)\right) - \frac{\beta}{2} I_{SHG}(r,z,t) B_{SHG}(r,z,t) \quad (1)$$

$$\frac{\partial B_p(r,z,t)}{\partial z} + i\mathcal{F}^{-1}\big(\beta(\omega + \omega_p) B_p(r,z,\omega)\big) = -i\chi(z)\mathcal{F}^{-1}\left(\frac{\omega + \omega_p}{n(\omega + \omega_p)c} \mathcal{F}\left(B_{SHG}(r,z,t) B_p^*(r,z,t)\right)\right) \quad (2)$$

Incorporation of the measured FROG pump pulse and the measured 40μm FWHM Gaussian beam profile to the simulation, yielded great agreement with experimental results for $\beta = 4 \left[\frac{cm}{GW}\right]$, in correspondence with the nonlinear coefficient obtained in [22,23].

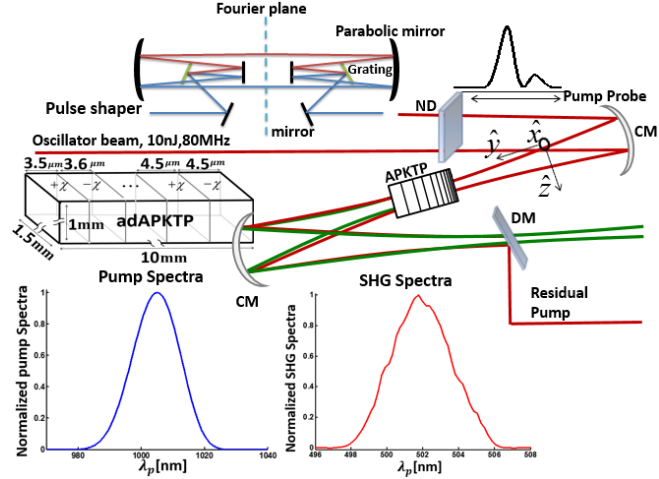

Fig. 1. (Color online) The adiabatic SHG experimental apparatus. The programmable 4-f of the quadratic phase tunable pulse shaper is composed of a pair of diffraction gratings with 600 lines/mm and a pair of parabolic curve mirror. Two-dimensional SLM was placed at the Fourier plane and was used as a dynamic filter for spectral phases. ND - neutral density filter, CM - curved mirror, DM- dichroic mirror, adAPKTP - adiabatically aperiodically poled KTP. Left lower inset - the measured pump normalized spectra. Right lower inset - the measured SHG normalized spectra.

The energy conversion efficiency of the crystal is defined to be the ratio between the generated SHG pulse energy to the pump pulse energy:

$$\eta = \frac{E_{SHG}(z_{final})}{E_p(z_{initial})} \quad (3)$$

Where $z_{final}$ and $z_{initial}$ are the locations of the crystal output and input facets, respectively. The conversion efficiency measurements as function of the central pump wavelength are presented as red dots in Fig. 2(a). Although conversion efficiency measurements were limited in the range of

970-1030nm due to source limitations, the adiabatic design is capable of an efficient frequency doubling within a bandwidth of 80nm, shown as blue solid line.

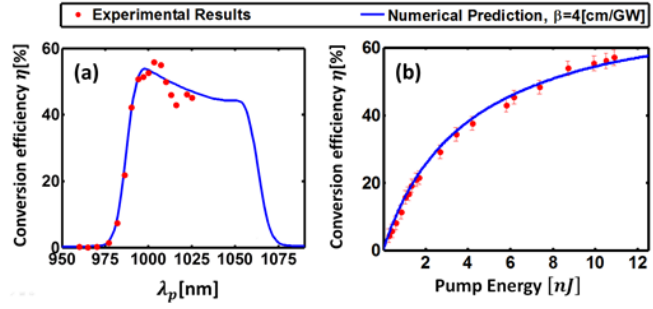

Fig. 2: Frequency conversion dependence with pump central wavelength and energy. (a) Conversion efficiency dependence with pump central wavelength. (b) Conversion efficiency as a function of pump energy centered at 1005nm wavelength. Deviation between the loss-free numerical predication [23] and new experimental results can be attributed to TPA.

Next, we have added a broadband 4-f pulse shaper based on computer controlled SLM positioned at its Fourier plane. First, by varying the spectral quadratic phase [24], we have used it as an aligned tunable compressor that can control of the GDD of the fundamental, which also influence SHG output spectral distribution. The architecture for the 4-f pulse shaper was includes two pieces of 600g/mm ruled grating which disperse the light angularity into a telescope based pair of 913.4 mm ROC parabolic mirrors, which fit especially for a 4f-arrangement or in a chirped pulse amplification system.

The pump probe apparatus were implemented by applying an absolute-valued spectral phase centered within the bandwidth of the fundamental spectral phase, denoted by $\omega_{Abs}$: $\phi \propto |\omega - \omega_{Abs}|$. Determined by the sign of the linear phase slope, different parts of the pump spectra are translated to opposite directions in the time domain, therefore enabling splitting the incoming pump pulse into two localized pump pulses, with extinction ratio and delay determined by $\omega_{Abs}$ and the absolute phase slope. The idler spectra dependence with $\lambda_{Abs} = \frac{2\pi c}{\omega_{Abs}}$ is presented in Fig. 3 for an input hyperbolic sech pump pulse centered on 1005nm. When applying the abs-valued shape spectral phase in the center of the pump pulse $\lambda_{Abs} = 1005nm$, as shown in Fig 3(a I), the initial pump pulse splits to two pulses with the same peak intensity, resulting in a symmetrical normalized pump probe SHG spectra. When the abs-valued shape phase is not symmetrical regarding the pump central frequency, as shown in Fig 3(a-II), the generated a-symmetrical pump pulses results in a-symmetrical SHG spectra.

We proceed by applying a π-step spectral phase. As a result of destructive interference between the different frequency components, the π-step induces a SHG/SFG pulses with a spectral dip at a different wavelengths. It worth noting, that due to the mathematical similarity between the perturbative solution of two photon absorption process in a non-resonant two level quantum systems and the instantaneous SHG process (where time scaling t is replaced by propagation length in crystal at the z axis is perform), most of the coherent control schemes that were applied in atomic physics community can be adopted in pulse-shaped SHG outcome using the ASHG nonlinear crystal [22]. In order to demonstrate shaping of the SHG spectra, we have also applied π-step function at different locations on the pump spectral phase.

We consider the case of second harmonic generation of ultrashort pulse with an electric field distribution of $\varepsilon(t)$. From second order non-resonant time-dependent perturbation theory, the instantaneous second harmonic spectral field can be found by the convolution of the fundamental spectral field $\tilde{\varepsilon}(\Omega) = F.T\{\varepsilon(t)\}$ [10], which describe in equation (4):

$$\varepsilon_{SHG}(\omega_0) = \left|\int_{-\infty}^{\infty} \epsilon^2(t) \cdot \exp(i\omega_0 t)\, dt\right|^2 = \left|\int_{-\infty}^{\infty} \tilde{\varepsilon}(\Omega)\tilde{\varepsilon}(\omega_0 - \Omega)d\Omega\right|^2 \qquad (4)$$

where $\omega_0$ is the SHG frequency.

If we perform change of variables as $\Omega \to \Omega + \frac{\omega_0}{2}$, and write explicitly $\tilde{\varepsilon}(\Omega) = A(\Omega)\exp[i\Phi(\Omega)]$, we obtain equation (5):

$$\varepsilon_{SHG}(\omega_0) = \left|\int_{-\infty}^{\infty} A\left(\frac{\omega_0}{2} + \Omega\right) A\left(\frac{\omega_0}{2} - \Omega\right) \cdot \exp\left[i\left\{\Phi\left(\frac{\omega_0}{2} + \Omega\right) + \Phi\left(\frac{\omega_0}{2} - \Omega\right)\right\}\right] d\Omega\right|^2 \qquad (5)$$

where $A(\omega)$ and $\Phi(\omega)$ are the spectral amplitude and spectral phase, respectively.

The equation reflects that the SHG/SFG occurs for all pairs of photons with frequencies which adds to $\omega_0$ and lie within the spectrum of the exciting pulse. It's easy to see that when the phase cancels (i.e. $\Phi\left(\frac{\omega_0}{2} + \Omega\right) + \Phi\left(\frac{\omega_0}{2} - \Omega\right) = 0$) and for a symmetric amplitude $A(\omega)$ in respect to $\frac{\omega_0}{2}$, the amount of SHG is maximized. This is in complete analogy to the finding of Meshulach et. al. [24, 25], where they showed that anti-symmetrical spectral phase can result in the same TPA rate as transform-limited excitation. The same behavior happens for the instantaneous SHG/SFG case, where the transform limit pulse ($\Phi(\Omega) = 0$) and every anti-symmetric phase compared to $\frac{\omega_0}{2}$ results in the same SHG rate, ie. The anti-symmetric phase can significantly affect the shape of the pulse to have much smaller amplitude and a much longer duration without affecting the SHG [23].

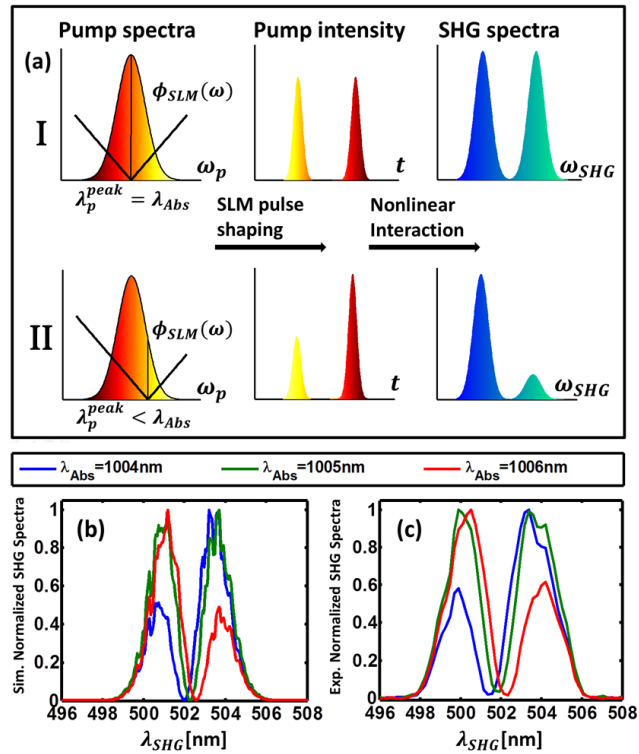

Fig 3: Absolute phase effects on the SHG pulse for different $\omega_{Abs}$, the central absolute phase frequency. (a) Illustration of absolute phase concept. (b) Simulated normalized SHG spectra. (c) Experimental normalized SHG spectra.

Our experimental results of applying an anti-symmetric of π-step phase each time at a different wavelength location at the SLM, show different SHG/SFG spectra for each measurement. As expected, that destructive interference will appear in the SHG/SFG if the step is not in the middle of the spectrum, and could even lead for complete destructive interference of the SHG/SFG signal at specific wavelength. The calculation can be generalized easily to any anti-symmetric phase. In Fig. 4, we show the experimental results as well as the simulations of broadband ASHG conversion when exciting the ASHG crystal by a pulse with varying spectral phase of π-step, as a function of the step position. The simulation results were obtained by inserting the measured spectrum of the original pulse. Very good agreement is obtained between the experimental results and the numerical simulations.

To conclude, we experimentally investigate the performance of an adiabatic aperiodically poled KTP crystal using conventional femtosecond high rep.-rate oscillator in the nJ energy level regime. It was shown that efficient wideband ultrashort second harmonic generation is feasible in a single crystal, experimentally demonstrating acceptance bandwidth bigger than 40nm with nJ level excitation. Furthermore, it was shown that the wideband operation of the adiabatic crystal design enables us to perform pulse shaping manipulations on the fundamental pulses, with no restriction imposed by the limited bandwidth of a conventional birefringence or periodic crystals designs.

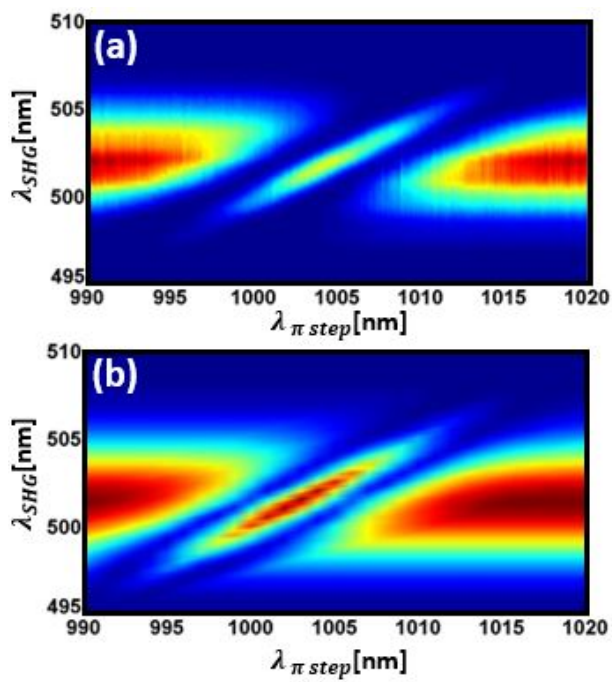

Fig.4: ASHG conversion in KTP crystal excited by a pulse with a spectral phase of π-step, as a function of the step position. (a) Experimental results. (b) Simulation results. The simulation results were obtained by inserting the measured spectrum of the original pulse.

SLM dependent tunable pump probe was given as an example, where altering the pump probe spectral and temporal characteristics is possible without the need of realigning the experimental apparatus. We show that the SHG/SFG spectrum can be manipulated by tailoring the shape of the exciting ultrashort pulse and the broad spectral conversion response of the ASHG nonlinear crystal. In particular, we investigate the effect of a spectral phase modulation of abs-value and a π -step spectral phase. We show that spectrum shaping as well as complete destructive interference appear in the SHG can be achieved. Also, we show that certain spectral phase modulation leads to long pulses that induce SHG that is effectively as transform-limited pulses. The basic principles presented here open a wide new area for theoretical and experimental work, as well as possible applications in nonlinear spectroscopy and in atomic and molecular physics.

This research has received funding from the European Research Council (ERC) under the European Union's Horizon 2020 research and innovation programme (grant No 639402)

## References

1. J.-C. Diels and W. Rudolph, "Ultrashort Laser Pulse Phenomena: Fundamentals, Techniques, and Applications on a Femtosecond Time Scale", Academic Press, 2nd Edition (2006).
2. H. A. Zewail, "Femtochemistry: Atomic-scale dynamics of the chemical bond using ultrafast lasers." Angewandte Chemie International Edition 39 2586-2631 (2000).
3. A. M. Weiner, "Ultrafast Optics", Wiley (2008).
4. L. A. Cavalieri, N. Müller, Th. Uphues, V. S. Yakovlev, A. Baltus caronka, B. Horvath, B. Schmidt, L. Blümel, R. Holzwarth, S. Hendel, M. Drescher, U. Kleineberg, P. M. Echenique, R. Kienberger, F. Krausz & U. Heinzmann, "Attosecond spectroscopy in condensed matter." Nature 449 1029 (2007).
5. A. V. Kimel, A. Kirilyuk, P. A. Usachev, R. V. Pisarev, A. M. Balbashov & Th. Rasing. "Ultrafast non-thermal control of magnetization by instantaneous photomagnetic pulses." Nature 435, 655-657 (2005).
6. N. Dudovich, D. Oron, and Y. Silberberg. "Quantum Coherent Control for Nonlinear Spectroscopy and Microscopy." Annual Rev. of Phys. Chem. 60, 277-292 (2009).
7. R. Trebino, "Frequency-resolved optical gating: the measurement of ultrashort laser pulses", Springer Science & Business Media (2012).
8. V.V Lozovoy, I. Pastirk, and M. Dantus, "Multiphoton intrapulse interference. IV. Ultrashort laser pulse spectral phase characterization and compensation." Opt. Letters 29, 775-777 (2004).
9. C. Iaconis, and I. A. Walmsley, "Spectral phase interferometry for direct electric-field reconstruction of ultrashort optical pulses." Opt. Letters 23, 792-794 (1998).
10. R.W. Boyd, "Nonlinear Optics". Burlington: 3rd Edition (2011).
11. H. Suchowski, B. Bruner, A. Arie and Y. Silberberg, "Broadband nonlinear frequency conversion." Opt. and Phot. News 21, 36-41 (2010).
12. H. Suchowski, G. Porat, and A. Arie, Laser and Phot. Rev., 8, 333–367 (2014).
13. C. Heese, C. R. Phillips, B. W. Mayer, L. Gallmann, M. M. Fejer, and U. Keller. "75 MW few-cycle mid-infrared pulses from a collinear apodized APPLN-based OPCPA." Opt. Express 20, 6888-26894 (2012).
14. H. Suchowski, V. Prabhudesai, D. Oron, A. Arie and Y. Silber berg, Opt. Express 17 12731–12740 (2009).
15. C. Heese, C. R. Phillips, L. Gallmann, M. M. Fejer and U. Keller. "Ultrabroadband, highly flexible amplifier for ultrashort midinfrared laser pulses based on aperiodically poled Mg:LiNb O$_3$" Opt. Letters 35, 2340–2342 (2010).

16. C. Heese, C.R. Phillips, L. Gallmann, M.M Fejer and U. Keller "Role of apodization in optical parametric amplifiers based on aperiodic quasi-phasematching gratings" Opt. Express 20, 18066–18071 (2012).
17. O. Yaakobi, L. Caspani, M. Clerici, F. Vidal and R. Morandotti. "Complete energy conversion by autoresonant three-wave mixing in nonuniform media" Opt. Express 21, 1623–1632 (2013).
18. H. Suchowski, P. R. Krogen, S. W. Huang, F. X. Kartner and J. Moses. "Octave-spanning coherent mid-IR generation via adiabatic difference frequency conversion" Opt. Express 21, 28892–28901 (2013).
19. P. Krogen, H. Suchowski, H. Liang, N. Flemens, K.-H. Hong, F. X. Kärtner, J. Moses, "Generation and Arbitrary Shaping of Intense Single-Cycle Pulses in the Mid-Infrared", Nat. Photonics, 222–226 (2017).
20. G. Porat and Ady Arie. "Efficient, broadband, and robust frequency conversion by fully nonlinear adiabatic three-wave mixing." JOSA B 30 1342-1351 (2013).
21. C. R. Phillips, C. Langrock, D. Chang, Y. W. Lin, L. Gallmann, M. M. Fejer , "Apodization of chirped quasi-phasematching devices", JOSA B 30, 1551-1568 (2013).
22. A. Leshem, G. Meshulam, G. Porat G and A. Arie "Adiabatic second-harmonic generation" Optics Letters, 41 1229–1232 (2016).
23. A. Dahan, A. Levanon, M. Katz and H. Suchowski "Ultrafast adiabatic second harmonic generation" J. Phys. Condens. Matter (2016).
24. V. A Maslov, V. A Mikhailov, O. P Shaunin and I A Shcherbakov "Nonlinear absorption in KTP crystals", 27, 356–359 (1997).
25. D. Meshulach, Y. Silberberg, "Coherent quantum control of two-photon transitions by a femtosecond laser pulse", Nature 396, 239-242 (1998).
26. D. Meshulach and Y. Silberberg, "Coherent quantum control of multiphoton transitions by shaped ultrashort optical pulses", Phys. Rev. A, 60, 1287 (1999).

## References

1. Jean-Claude Diels and Wolfgang Rudolph, "Ultrashort Laser Pulse Phenomena: Fundamentals, Techniques, and Applications on a Femtosecond Time Scale" (Second Edition).
2. Ahmed H. Zewail. "Femtochemistry: Atomic-scale dynamics of the chemical bond using ultrafast lasers (Nobel Lecture)." Angewandte Chemie International Edition 39.15 (2000): 2586-2631.
3. Andreo M. Weiner, "Ultrafast Optics", Wiley (2008).
4. A. L. Cavalieri, N. Müller, Th. Uphues, V. S. Yakovlev, A. Baltus caronka, B. Horvath, B. Schmidt, L. Blümel, R. Holzwarth, S. Hendel, M. Drescher, U. Kleineberg, P. M. Echenique, R. Kienberger, F. Krausz & U. Heinzmann, "Attosecond spectroscopy in condensed matter." Nature 449 1029-1032 (2007).
5. A. V. Kimel, A. Kirilyuk, P. A. Usachev, R. V. Pisarev, A. M. Balbashov & Th. Rasing. "Ultrafast non-thermal control of magnetization by instantaneous photomagnetic pulses." Nature 435, 655-657 (2005).
6. Dudovich, Nirit, Dan Oron, and Yaron Silberberg. "Single-pulse coherently controlled nonlinear Raman spectroscopy and microscopy." Nature 418.6897 (2002): 512-514.
7. Trebino, Rick. Frequency-resolved optical gating: the measurement of ultrashort laser pulses. Springer Science & Business Media, 2012.
8. Lozovoy, Vadim V., Igor Pastirk, and Marcos Dantus. "Multiphoton intrapulse interference. IV. Ultrashort laser pulse spectral phase characterization and compensation." Optics letters 29.7 (2004): 775-777.
9. Iaconis, Chris, and Ian A. Walmsley. "Spectral phase interferometry for direct electric-field reconstruction of ultrashort optical pulses." Optics letters 23.10 (1998): 792-794.
10. Boyd RW2008 Nonlinear Optics (Third Edition) ed BoydR W (Burlington: Academic Press) third edition ed ISBN 978-0-12-369470-6.
11. Haim Suchowski, Barry D. Bruner, Ady Arie, and Yaron Silberberg. "Broadband nonlinear frequency conversion." Optics and Photonics News 21.10 (2010): 36-41.
12. H. Suchowski, Gil Porat, and Adi Arie, Laser and Phot. Rev., 8, 333–367 (2014).
13. C. Heese, C. R. Phillips, B. W. Mayer, L. Gallmann, M. M. Fejer, and U. Keller. "75 MW few-cycle mid-infrared pulses from a collinear apodized APPLN-based OPCPA." Optics express 20.24 (2012): 26888-26894.
14. Haim Suchowski, Vaibhav Prabhudesai, Dan Oron, Ady Arie, and Yaron Silberberg. "Robust adiabatic sum frequency conversion" Opt. Express 17 12731–12740 (2009).
15. C. Heese, C. R. Phillips, L. Gallmann, M. M. Fejer and U. Keller, "Ultrabroadband, highly flexible amplifier for ultrashort midinfrared laser pulses based on aperiodically poled Mg:LiNbO$_3$" Opt. Lett. 35 2340–2342 (2010).
16. C. Heese, C.R. Phillips, L. Gallmann, M.M Fejer and U. Keller. "Role of apodization in optical parametric amplifiers based on aperiodic quasi-phasematching gratings" Express 20 18066–18071 (2012).
17. O. Yaakobi, L. Caspani, M. Clerici, F. Vidal and R. Morandotti. "Complete energy conversion by autoresonant three-wave mixing in nonuniform media" Opt. Express 21 1623–1632 (2013).
18. Haim Suchowski, Peter R. Krogen, Shu-Wei Huang, Franz X. Kärtner, and Jeffrey Moses. "Octave-spanning coherent mid-IR generation via adiabatic difference frequency conversion" 2013 Optics express 21 28892–28901
19. Peter Krogen, Haim Suchowski, Houkun Liang, Noah Flemens, Kyung-Han Hong, Franz X. Kärtner & Jeffrey Moses. "Generation and Arbitrary Shaping of Intense Single-Cycle Pulses in the Mid-Infrared", Nat. Photonics, 222–226 (2017).
20. Porat Gil, and Ady Arie. "Efficient, broadband, and robust frequency conversion by fully nonlinear adiabatic three-wave mixing." JOSA B 30.5 (2013): 1342-1351.
21. C. R. Phillips, C. Langrock, D. Chang, Y. W. Lin, L. Gallmann, M. M. Fejer , "Apodization of chirped quasi-phasematching devices", JOSA B, 30, 1551-1568 (2013).
22. Anat Leshem, Guilia Meshulam, Gil Porat, and Ady Arie. "Adiabatic second-harmonic generation" Optics Letters, 41 1229–1232 (2016).
23. Asaf Dahan, Assaf Levanon, Moti Katz and Haim Suchowski "Ultrafast adiabatic second harmonic generation" J. Phys. Condens. Matter (2016).
24. Vladislav A Maslov, V A Mikhailov, O P Shaunin and Ivan A Shcherbakov."Nonlinear absorption in KTP crystals", 27, 356–359 (1997).
25. Doron Meshulach & Yaron Silberberg. "Coherent quantum control of two-photon transitions by a femtosecond laser pulse", Nature 396, 239-242 (1998).
26. Doron Meshulach & Yaron Silberberg. "Coherent quantum control of multiphoton transitions by shaped ultrashort optical pulses", Phys. Rev. A, 60, 1287 (1999).